\documentclass[letterpaper, 10 pt, conference]{ieeeconf}
\IEEEoverridecommandlockouts 
\usepackage{cite}
\usepackage{amsmath,amssymb,amsfonts}
\usepackage{graphicx}
\usepackage{caption}
\usepackage{floatrow}
\usepackage{subcaption}
\usepackage{textcomp}
\usepackage{amsfonts, amsmath}
\usepackage{amssymb}
\usepackage{cleveref}
\usepackage{amsmath}
\usepackage{graphicx}
\usepackage{multirow}
\usepackage{etoolbox}
\makeatletter
\patchcmd{\@makecaption}
  {\scshape}
  {}
  {}
  {}
\makeatother

\newtheorem{theorem}{Theorem}
\newtheorem{lemma}{Lemma}

\newtheorem{remark}{Remark}
\newtheorem{assumption}{Assumption}
\newtheorem{property}{Property}
\usepackage{cuted}
\usepackage[english]{babel}
\usepackage{xcolor}
    \makeatletter
 \let\old@ps@headings\ps@headings
 \let\old@ps@IEEEtitlepagestyle\ps@IEEEtitlepagestyle
 \def\confheader#1{
 \def\ps@headings{
 \old@ps@headings
 \def\@oddhead{\strut\hfill#1\hfill\strut}
 \def\@evenhead{\strut\hfill#1\hfill\strut}
 }
 \def\ps@IEEEtitlepagestyle{
 \old@ps@IEEEtitlepagestyle
 \def\@oddhead{\strut\hfill#1\hfill\strut}
 \def\@evenhead{\strut\hfill#1\hfill\strut}
 }
 \ps@headings
 }
 \makeatother
 \usepackage[pscoord]{eso-pic}

  \makeatletter
\newcommand{\linebreakand}{%
  \end{@IEEEauthorhalign}
  \hfill\mbox{}\par
  \mbox{}\hfill\begin{@IEEEauthorhalign}
}
\makeatother  
\begin{document}
\title{State and Input Constrained Adaptive Tracking Control of Uncertain Euler-Lagrange Systems with Robustness and Feasibility Analysis
}
\author{Poulomee~Ghosh and Shubhendu~Bhasin
\thanks{Poulomee Ghosh and Shubhendu Bhasin are with Department of Electrical Engineering, Indian Institute of Technology Delhi, New Delhi, India. 
       {\tt\small (Email: Poulomee.Ghosh@ee.iitd.ac.in, sbhasin@ee.iitd.ac.in)}}}
\maketitle
\begin{abstract}
This paper proposes an adaptive tracking controller for uncertain Euler-Lagrange (E-L) systems with user-defined state and input constraints in presence of bounded external disturbances. A barrier Lyapunov function (BLF) is employed for state constraint satisfaction, integrated with a saturated controller that ensures the control input remains within pre-specified bounds. To the best of the authors' knowledge, this is the first result on tracking control of state and input-constrained uncertain E-L systems that provides verifiable conditions for the existence of a feasible control policy. The efficacy of the proposed controller in terms of constraint satisfaction and tracking performance is demonstrated through simulation on a robotic manipulator system. 

\end{abstract}

\section{Introduction}
\label{sec:intro}
Control of nonlinear systems, especially Euler-Lagrange (E-L) systems, is fundamental to modern control theory due to their wide applications in mechanical, robotic, and aerospace domains. In many practical scenarios, these systems are often subject to various constraints, including physical, safety, or performance limitations, which can be translated into bounds on plant states and control input. 
Designing controllers for systems with uncertainties and constraints is challenging. Conventional adaptive and robust control techniques \cite{arc1, arc2} are primarily designed for systems with parametric uncertainties and external disturbances, and typically guarantee the boundedness of both plant states and control inputs. However, these bounds are neither known a priori nor user-defined. It is essential to ensure the stability, robustness, and feasibility of the system in the presence of constraints. Therefore, it is crucial and practically relevant to develop a robust adaptive controller that provides a feasible control law to ensure tracking while satisfying user-defined state and input constraints.\\
Several approaches have been developed to explicitly address state constraints in control systems, including model predictive control (MPC) \cite{mpc1},\cite{mpc}, control barrier function (CBF)-based approach \cite{CBF2}, \cite{CBF3}, optimal control theory \cite{opt1},\cite{opt}, invariant set theory \cite{blanchini},\cite{set}, reference governor approach \cite{rga}, \cite{gilbert} etc. However, most of these methods involve running a computationally intensive optimization routine at each time step, especially for systems with fast dynamics or limited computational power. Moreover, most of the approaches rely on complete model knowledge.\\
An alternative safety certificate, the Barrier Lyapunov Function (BLF) \cite{BLF},\cite{BLF2}, has been extensively used to ensure state constraints by incorporating an error transformation \cite{elblf1}. Despite its effectiveness, existing BLF-based approaches often lead to excessive control effort when the states approach the boundary of the constrained region, risking violations of the actuator’s limits. To address this, imposing user-defined bounds on the control effort, alongside state constraints, is essential for extending the applicability of these methods to safety-critical systems.\\
For input-constrained systems, multiple techniques, including saturated functions \cite{incon10, inconnew, incon11, incon12, incon13}, saturated feedback controllers \cite{annaswamy}, \cite{Lav}, reference governors \cite{garone2017reference} etc., have been employed. However, the simultaneous satisfaction of constraints on both state and input remains a relatively less explored area of research. Although limited, a few techniques tackle the tracking control problem for uncertain nonlinear systems subject to both state and input constraints. MPC \cite{dhar, mpc11, dhar2} is a well-established approach that incorporates both types of constraints in an optimization routine; however, the computational complexity increases multifold with state dimension and prediction horizon. In applications with strict real-time requirements, the optimization overhead of MPC may be undesirable. Additionally, MPC typically assumes a known system model but may face challenges with parametric uncertainties unless combined with robust or adaptive strategies \cite{robustmpc1}, which can significantly increase the complexity of the control design. In \cite{zcbf1}, a zeroing control barrier function (ZCBF) is introduced for E-L systems to enforce both state and input constraints. However, the approach relies on complete model knowledge and requires solving a Quadratic Program (QP) in real-time to ensure constraint satisfaction, which significantly increases computational load and poses implementation challenges for systems with limited processing power. Moreover, since CBF-based methods separate constraint satisfaction from stability analysis, they often result in a more complex design process and may lead to suboptimal performance. \\
In our previous work \cite{myel1}, an adaptive tracking controller was developed for an uncertain EL system that deals with user-defined constraints on both state and input. However, the design was based on the restrictive assumption that the required control effort is within the saturation limit for all time instances, and no feasibility condition was provided. Moreover, the design considered only a single constraint on the norm of the plant states rather than imposing separate constraints for individual states and required complete knowledge of the mass matrix.\\
The primary contribution of this paper is to provide verifiable conditions for the existence of a feasible control policy that ensures tracking to the desired reference trajectory while adhering to user-defined state and input constraints in the presence of parametric uncertainties and bounded external disturbances.
A saturated feedback controller, which is designed to tackle the input constraint, is strategically combined with a BLF-based controller that ensures state constraint satisfaction. The robustness of the controller is guaranteed by projection-based adaptive update laws. 
Unlike MPC and CBF-based approaches, which typically depend on real-time optimization and precise model knowledge, the proposed method is computationally efficient, directly incorporates constraints into the Lyapunov framework, and avoids optimization entirely. It provides explicit bounds on states and inputs while integrating with an adaptive controller to handle parametric uncertainties. The inclusion of a projection-based adaptive control law adds robustness against bounded external disturbances, making it applicable for real-time control of uncertain E-L systems.
To the authors' best knowledge, this is the first result in adaptive control literature, which is optimization-free and ensures the robustness and feasibility of the control policy while simultaneously addressing user-defined constraints on both state and input.\\
\textit{Notation:} In this paper, $\mathbb{R}$ represents the set of real numbers, $\mathbb{R}^{p \times q}$ denotes the set of $p\times q$ real matrices, and $I_{p}$ refers to the identity matrix in $\mathbb{R}^{p \times p}$. $\log(\cdot)$ denotes the natural logarithm of $(\cdot)$, and $\|.\|$ represents the Euclidean vector norm and its induced matrix norm. $\zeta^{(i)}(t)$ signifies the $i^{th}$ derivative of $\zeta$ with respect to time. The trace of $A\in \mathbb{R}^{n \times n}$ is represented by, $\text{tr}(A)$ and the eigenvalues of A with minimum and maximum real parts are denoted by $\lambda_{min}\{A\}$ and $\lambda_{max}\{A\}$, respectively.

\section{Problem Formulation}
\label{sec:LFSR}
Consider an Euler-Lagrange (E-L) system 
\begin{align}
    M(q)\ddot{q}+V_m(q,\dot{q})\dot{q}+G_r(q)+F_d(\dot{q})=\tau+d
    \label{plant11}
\end{align}
where $q(t), \dot{q}(t), \ddot{q}(t) \in \mathbb{R}^n$ denote the generalized position, velocity, and acceleration, respectively, $M(q)\in\mathbb{R}^{n\times n}$ denotes a generalized inertia matrix; $V_m(q,\dot{q})\in \mathbb{R}^{n\times n}$ denote an uncertain generalized centripetal-Coriolis matrix;  $G_r(q)\in \mathbb{R}^n$ and $F_d(\dot{q})\in \mathbb{R}^n$ represent the uncertain generalized gravity and friction vectors, respectively. The control input is denoted by $\tau=[\tau_1,\hdots,\tau_n]^T\in\mathbb{R}^n$ and the external disturbance is denoted by $d(t)\in\mathbb{R}^n$, which is assumed bounded, i.e., $\|d(t)\|\leq \bar{d}$, where $\bar{d}>0$ is a known constant.\\
The following standard properties \cite{Vidyasagar} of E-L dynamics are used for the subsequent development of control law and stability analysis.
\begin{property}
The inertia matrix $M(q)\in \mathbb{R}^{n \times n}$ is symmetric, positive-definite and satisfies the following inequality,
\begin{align}
    m_1\|\mu\|^2\leq \mu^TM(q)\mu \leq m_2\|\mu\|^2,
\end{align}
 where $m_1$ and $m_2$ are positive constants and $\mu \in \mathbb{R}^n$ is an arbitrary vector.
\end{property}

\begin{property}
 The matrix $\dot{M}(q)-2V_m(q,\dot{q})$ is skew-symmetric, i.e.,
\begin{align}
   \mu^T(\dot{M}(q)-2V_m(q,\dot{q}))\mu=0, && \forall \mu\in\mathbb{R}^n
\end{align}
\end{property}

\begin{property}
The E-L dynamics is considered to be linearly parameterizable,
\begin{align}
    Y(q,\dot{q},\ddot{q})\theta=M(q)\ddot{q}+V_m(q,\dot{q})\dot{q}+G_r(q)+F_d(\dot{q})
    \label{Property4}
\end{align}
where $Y:\mathbb{R}^n\times \mathbb{R}^n\times \mathbb{R}^{n} \rightarrow \mathbb{R}^{n\times m}$ is the known regression matrix and $\theta\in \mathbb{R}^m$ is the unknown parameter vector.
\end{property}
The control objective is to design a feasible control policy $\tau(t)$ for the uncertain system (\ref{plant11}) such that the plant states $q(t)$, $\dot{q}(t)$ track the user-defined reference trajectory $q_d(t)$, $\dot{q}_d(t)$, respectively, as $t\rightarrow \infty$ while simultaneously satisfying the following constraints on the plant states and control input.\\~\\
\textbf{State Constraint:} The plant states are uniformly bounded within a pre-defined safe set given by $\Omega_q=\{q(t),\dot{q}(t) \in \mathbb{R}^n: \|q(t)\|< \bar{Q}, \|\dot{q}(t)\|<\bar{\mathcal{V}} \}$, where $\bar{Q},\bar{\mathcal{V}}$ are user-defined positive constants.\\~\\
\textbf{Input Constraint:} The control input is uniformly bounded within a pre-specified safe set given by $\Omega_{\tau}=\{\tau(t)\in \mathbb{R}^n: \|\tau(t)\|\leq \bar{\tau} \}$, where $\bar{\tau}$ is a user-defined positive constant.

\section{Proposed Methodology}
The position and velocity tracking errors are defined as $e(t)\triangleq q(t)-q_d(t)$ and $\dot{e}(t)\triangleq \dot{q}(t)-\dot{q}_d(t)$, respectively.
To facilitate the design, a filtered  tracking error $r(t)\in \mathbb{R}^n$ is defined as
\begin{align}
    &r=\dot{e}+\alpha e
    \label{fte}
\end{align}
where $\alpha\in \mathbb{R}$ is a positive constant. Differentiating (\ref{fte}) and using (\ref{Property4}), the dynamics in (\ref{plant11}) can be written as
\begin{align}
  M\dot{r}=Y\theta+\tau-V_mr+d 
\end{align}
where, $Y\in\mathbb{R}^{n\times m}$ is the known regressor matrix, $\theta\in\mathbb{R}^{m}$ is unknown parameter vector. Here, $Y\theta$ is given by
\begin{align}
    Y\theta=M(\alpha \dot{e}-\ddot{q}_d)+V_m(r-\dot{q})-F_d-G_r
\end{align}

\subsection{Input Constraint Satisfaction Using Saturated Control Design}

An auxiliary control input $u(t)\in \mathbb{R}^n$ is considered as
\begin{align}
&u=-Y\hat{\theta}-K_1r
\label{pc1}
\end{align}
where $u(t)\triangleq[u_1(t),\hdots, u_n(t)]^T $, $\hat{\theta}(t)\in\mathbb{R}^m$ is the estimated unknown parameter vector and $K_1\in \mathbb{R}^{n \times n}$ is a positive controller parameter gain. A saturated feedback controller is designed as
\begin{align}
&\tau_i(t)= \begin{cases}
u_i(t) & \text{if}\:\:\: \|u(t)\|\leq \bar{\tau}\\
\frac{\bar{\tau}}{\|u(t)\|}u_i(t) & \text{if}\:\:\: \|u(t)\|>\bar{\tau}
\end{cases},
\hspace{4pt} i=1, \hdots , n
\label{pc2}
\end{align}
Using (\ref{pc1}) and (\ref{pc2}), the closed-loop dynamics of the filtered tracking error can be expressed as
\begin{align}
    M\dot{r}=Y\tilde{\theta}-K_1r-V_mr+\Delta \tau+d 
    \label{edot}
\end{align}
where $\tilde{\theta}(t)\triangleq\theta(t)-\hat{\theta}(t)\in \mathbb{R}^{n}$ is the parameter estimation error and $\Delta \tau(t)\in\mathbb{R}^{m}$ is defined as the difference between the actual control input $\tau(t)$ and auxiliary control input $u(t)$, i.e., $\Delta \tau(t) \triangleq \tau(t)-u(t)$. Due to input saturation, an additional term \(\Delta \tau\) appears in the closed-loop error dynamics (\ref{edot}), which can be considered a disturbance-like term that the controller is expected to reject.

\begin{assumption}
 For state-constraint bounds $\bar{Q},\bar{\mathcal{V}}>0$, there exist known positive constants $\bar{Q}_d$, $\bar{\mathcal{V}}_d$ and $\alpha_3$ such that the desired reference trajectory $q_d(t)\in \mathbb{R}^n$ and its derivatives $\dot{q}_d(t),\ddot{q}_d(t)$ are bounded as
\begin{align}
    & \|q_d(t)\| \leq \bar{Q}_d < \bar{Q} && \forall t \geq 0 \label{assump1a}\\
    & \|\dot{q}_d(t)\|\leq \bar{\mathcal{V}}_d < \bar{\mathcal{V}} && \forall t \geq 0 \label{assump1b}\\
    & \|\ddot{q}_d(t)\|\leq \alpha_3 && \forall t \geq 0 \label{assump1c}
    \end{align}
\end{assumption}
Provided Assumption 1 holds, the state constraints can be transformed to constraints on the tracking error states: $\|e(t)\|<\mathcal{E}_Q$, $\|\dot{e}(t)\|<\mathcal{E}_{\mathcal{V}}$, $\forall t\geq 0$, where $\mathcal{E}_{Q},\mathcal{E}_{\mathcal{V}} \in \mathbb{R}$ are positive constants given by $\mathcal{E}_{Q}=\bar{Q}-\bar{Q}_d$ and $\mathcal{E}_{\mathcal{V}}=\bar{\mathcal{V}}-\bar{\mathcal{V}}_d$, i.e. $\|e(t)\|<\mathcal{E}_{Q} \implies\|q(t)\|\leq\bar{Q}$, and $\|\dot{e}(t)\|<\mathcal{E}_{\mathcal{V}} \implies\|\dot{q}(t)\|\leq\bar{\mathcal{V}}$.
\begin{assumption}
The initial condition of the reference model states $q_d(0), \dot{q}_d(0)$ are chosen such that the initial trajectory tracking error and filtered tracking error satisfy 
\begin{align}
   &\|e(0)\|\leq \mathcal{E}_{Q}-\frac{\kappa}{\alpha} \label{ezero}\\
   &\|{r}(0)\|<\kappa 
\end{align}
where, $\kappa=\mathcal{E}_{\mathcal{V}}-\alpha\mathcal{E}_Q$ is a user-defined positive constant.
\label{error_assumption}
\end{assumption}
\begin{remark}
    Assumption 2 implies that the plant states initially remain within the user-defined safe set to prevent constraint violation at the initial time, a standard assumption in constrained control design \cite{BLF, BLF2, elblf1}. Further, in $(\ref{ezero})$, the additional constraint on $\|e(0)\|$  facilitates the conversion of the constraint from trajectory tracking error to filter tracking error, thereby simplifying the subsequent analysis.
\end{remark}
\subsection{State Constraint Satisfaction using BLF}\label{AA}
To maintain the system states within the user-defined bounds, a BLF-based approach \cite{BLF} is employed. The following lemma establishes a basic result on constraint satisfaction with a BLF-based control strategy, ensuring both the states and the associated tracking errors remain bounded when the BLF is applied.

\begin{lemma}
For any positive constant $\kappa$, let $\Omega_r := \{r \in \mathbb{R}^n : \|r\|<\kappa\}\subset \mathbb{R}^n$ and $\Psi:=\mathbb{R}^N\times \Omega_r \subset \mathbb{R}^{N+n}$ be open sets. Consider the system dynamics given by
\begin{align}
    \dot{\mu}=f(t,\mu)
\end{align}
$\mu:=[r^T,\xi^T]^T\in \Psi$, where $\xi$ is the augmentation of the unconstrained states and the function $f:\mathbb{R}_{+}\times \Psi \rightarrow \mathbb{R}^{N+n}$ is measurable  for each fixed $\mu$ and locally Lipschitz in $\mu$, piecewise continuous and locally integrable on $t$. Suppose there exists a positive definite, decrescent, quadratic candidate Lyapunov function $V_2(\xi):\mathbb{R}^N\rightarrow\mathbb{R}_{+}$ and a continuously differentiable, positive-definite, scalar function $V_1(r):\Omega_r \rightarrow \mathbb{R}_{+}$, defined in an open region containing the origin such that 
\begin{align}
    V_1(r)\rightarrow \infty \hspace{10pt} \text{as} \hspace{10pt} \|r(t)\|\rightarrow \kappa
\end{align}
The candidate Lyapunov function can be written as $V(\mu)=V_1(r)+V_2(\xi)$. Provided $r(0)\in\Omega_r$, if the following inequality holds,
\begin{align}
    \dot{V}=\frac{\partial V}{\partial \mu}f< 0
\end{align}
 then $r(t)\in \Omega_r$  and $\xi(t)$ remain bounded $\forall t\geq 0$. \\
\begin{proof}
For the proof of Lemma 1, see \cite{BLF}.
\end{proof}

\end{lemma}
Using Property 1, $r^TMr<\bar{m}\|r\|^2$, and $\bar{m}$ is assumed to be known. To constrain $r(t)$, which in turn ensures error and state constraint satisfaction, we consider a BLF $V_1(r)$ defined on the set $\Omega_r^{'}:\{r\in\mathbb{R}:\bar{m}\|r\|^2<\kappa_m^2\}$, such that
\begin{align}
    &V_1(r)\triangleq\frac{1}{2}\log \frac{\kappa_m^2}{\kappa_{m}^{2}-\bar{m}\|r\|^2}
\end{align}
where $\kappa_m=\kappa\sqrt{\bar{m}}$. If $\bar{m}\|r\|^2\rightarrow \kappa_m^{2}$, i.e.,
when the constrained state $r(t)$ approaches the boundary of the safe set, the BLF $V_1(r)\rightarrow \infty$.
\begin{assumption}
    The norm of the unknown controller parameter vector is bounded such that $\|\theta\|<\bar{\theta}$, where $\bar{\theta}>0$ is assumed to be known.
\end{assumption}
\begin{remark}
    Assumption 3 is commonly used in projection-based adaptive control literature \cite{arc1}, where adaptive parameters are bounded within a predefined set, avoiding parameter drift in the presence of disturbances. In practical systems, unknown parameters are often naturally bounded due to physical constraints or modeling assumptions, and therefore, it is reasonable to assume the bound  \(\bar{\theta}\). Moreover, by incorporating system-specific information, \(\bar{\theta}\) can be estimated without being overly conservative. 
\end{remark}
For the proposed controller (\ref{pc1}) and (\ref{pc2}), the following adaptive update laws are defined, which will be employed in the subsequent Lyapunov analysis.
\begin{align}
    \dot{\hat{\theta}}=\text{proj}_{\Omega_{\theta}}\bigg(\frac{\Gamma Y^Tr}{\kappa_{m}^{2}-\bar{m}\|r\|^2}\bigg)
    \label{updatelaw}
\end{align}
where $\Gamma\in\mathbb{R}^{m\times m}$ is a positive-definite matrix. The projection operator \cite{lavretsky2012robust}, denoted as \(\text{proj}(\cdot)\), ensures that parameters remain bounded within a convex and compact region in the parameter space defined by \(\Omega_{\theta}=\{\hat{\theta}\in\mathbb{R}^m|\|\hat{\theta}\|^2 \leq \bar{\theta}^2\}\). 

\begin{theorem}
For the E-L system (\ref{plant11}), provided Assumptions 1-3 hold and the following feasibility condition (C1) is satisfied:\\
\textbf{C1:} The bound on the control input satisfies the following inequality,
\begin{align}
    \bar{\tau}>\omega_1+\omega_2\bar{\mathcal{V}}-\omega_3\bar{Q}
    \label{f1}
\end{align}
where $\omega_1=\bar{\theta}(\bar{\mathcal{V}}_d +\alpha_3+2)+\bar{d}- \omega_2\bar{\mathcal{V}}_d+\omega_3 \bar{Q}_d$, \(\omega_2=\lambda_{max}\{K_1\}+\bar{\theta}(2\alpha+3)-\lambda_{min}\{K_1\})\) and $\omega_3=\alpha(\omega_2-\bar{\theta}(\alpha+1))$ are known positive constants and \(\alpha\in \mathbb{R}\) satisfies the following gain condition\\
\begin{align}
\alpha<\frac{\mathcal{E}_{\mathcal{V}}}{\mathcal{E}_Q}
  \label{alpha}
\end{align} 
where $\mathcal{E}_{\mathcal{V}}$, $\mathcal{E}_Q$ are defined in Section II, the proposed controller (\ref{pc1}), (\ref{pc2}) and the adaptive update law (\ref{updatelaw}) ensure the following:
\begin{enumerate}
    \item [(i)] The plant states remain within the user-defined safe region, i.e. $\|q(t)\|< \bar{Q}, \|\dot{q}(t)\|< \bar{\mathcal{V}}$ $\forall t \geq 0$.
    \item [(ii)] The control effort is limited within the pre-specified bound, i.e. $\|\tau(t)\|\leq \bar{\tau}$ $\forall t \geq 0$.
    \item [(iii)] All the closed-loop signals remain bounded.
\end{enumerate} 
\end{theorem}

\begin{proof}
Consider the candidate Lyapunov function $V(\mu):\Omega_r\times \mathbb{R}^{N}\rightarrow \mathbb{R}_{+}$ as,
\begin{align}
    V(r,\tilde{\theta})&=
    \frac{1}{2}
    \bigg[\log\frac{\kappa_m^2}{\kappa_{m}^{2}-\bar{m}\|r\|^2}+\tilde{\theta}^T\Gamma^{-1}\tilde{\theta}\bigg]
    \label{lyap}
\end{align}
Taking the time-derivative of $V$ along the system trajectory 
\begin{align}
    \dot{V}=&\frac{1}{2(\kappa_{m}^{2}-\bar{m}\|r\|^2)}\bigg[2r^T(Y\tilde{\theta}-K_1r-V_mr+\Delta\tau+d)\nonumber\\
    &+r^T\dot{M}r\bigg]-\tilde{\theta}^T\Gamma^{-1}\dot{\hat{\theta}}
    \label{vdot111}
    \end{align}
    Employing Property 2 and adaptive update law (\ref{updatelaw}),
    
\begin{align}
    \dot{V}\leq&-\frac{r^TK_1r}{\kappa_{m}^{2}-\bar{m}\|r\|^2}+\frac{r^T\Delta \tau}{\kappa_{m}^{2}-\bar{m}\|r\|^2}+\frac{r^T\bar{d}}{\kappa_{m}^{2}-\bar{m}\|r\|^2} 
    \label{lyapfunc}
\end{align}
We now consider two cases depending on whether the controller is operating within the saturation limits (\textit{Case 1}) or is saturated (\textit{Case 2}).  \\

\textit{Case 1:} $\|u(t)\|\leq \bar{\tau},$ $i=1, \dots, n$\\
For this case, $\tau_i(t)=u_i(t)$ which implies $\|\tau (t)\|\leq\bar{\tau}$ and $\|\Delta \tau(t)\|=0$ $\forall t\geq 0$.\\

\textit{Case 2:} $\|u(t)\|> \bar{\tau},$ $i=1, \dots, n$\\
For this case, $\tau_i(t)=u_i(t)\frac{\bar{\tau}}{\|u(t)\|}$  and $\Delta\tau_i(t)=u_i(t)\frac{\bar{\tau}}{\|u(t)\|}-u_i(t)$ which bounds
\begin{align}
    &\|\Delta \tau(t)\|=\|u(t)\|(1-\frac{\bar{\tau}}{\|u(t)\|})
        \label{tau_bound}
    \end{align}
From (\ref{pc1}) the bound on \(\|u(t)\|\) can be calculated as
\begin{align}
    \|u\| 
    &\leq \|Y \hat{\theta}\|+\|K_1r\|\nonumber\\
    &\leq \bar{\theta}(\alpha\|\dot{e}\|+\|\ddot{q}_d\|+\|r\|+\|\dot{q}\|+2)+\lambda_{max}\{K_1\}\|r\|
    \label{vbound}
\end{align}
Now solving the differential equation (\ref{fte}) and employing Assumption \ref{error_assumption}, we can obtain  $\|e(t)\|\leq \|e(0)\|+\frac{\|r(t)\|}{\alpha}$ and $\|\dot{e}(t)\|\leq \alpha\|e(0)\|+2\|r(t)\|$ $\forall t\geq 0$, which in turn implies $\|\dot{q}(t)\|\leq \alpha \|e(0)\|+2\|r(t)\|+\bar{\mathcal{V}}_d$ $\forall t \geq 0$. substituting the bounds in (\ref{vbound}),
\begin{align}
    \|u\| 
    &\leq \Psi\|r\|+\xi
\end{align}
where $\Psi=\bar{\theta}(2\alpha +3)+\lambda_{max}\{K_1\}$ and $\xi=\bar{\theta}(\alpha^2\mathcal{E}_{Q}+\alpha\mathcal{E}_{Q}-\alpha\kappa-\kappa+\bar{\mathcal{V}}_d+\alpha_3+2)$ are positive constants. Substituting the bound of $\|u(t)\| $ in (\ref{tau_bound}),
\begin{align}
    \|\Delta \tau(t)\| < \Psi\|r\|+\xi-\bar{\tau}
    \label{tau_bound1}
\end{align}
Employing (\ref{tau_bound1}) in (\ref{lyapfunc}),
\begin{align}
    \dot{V}< &-\frac{\lambda_{min}\{K_1\}\|r\|^2}{\kappa_{m}^{2}-\bar{m}\|r\|^2}+\frac{\|r\|\bar{d}}{\kappa_{m}^{2}-\bar{m}\|r\|^2} \nonumber\\
    & +\frac{ (\Psi\|r\|+\xi-\bar{\tau})\|r\|}{\kappa_{m}^{2}-\bar{m}\|r\|^2}
    \\
    < &-\frac{\lambda_{min}\{K_1\}\|r\|}{\kappa_{m}^{2}-\bar{m}\|r\|^2}\bigg [\frac{ \bar{\tau}-\xi-\bar{d}}{\lambda_{min}\{K_1\}}\nonumber\\
     &-\bigg(\frac{\Psi}{\lambda_{min}\{K_1\}}-1\bigg)\|r\|\bigg]
    \label{vdot112}
\end{align}
Provided \(\bar{m}\|r(0)\|^2<\kappa_m^2\) at \(t=0\) which implies \(\|r(0)\|<\kappa\), to achieve a bounded result from (\ref{vdot112}), the following condition must hold.
\begin{align}
    \kappa<&\frac{\bar{\tau}-(\xi+\bar{d})}{\Psi-\lambda_{min}\{K_1\}}
    \label{c11}
\end{align}
substituting the value of $\Psi=\bar{\theta}(2\alpha +3)+\lambda_{max}\{K_1\}$, $\xi=\bar{\theta}(\alpha^2\mathcal{E}_{Q}+\alpha\mathcal{E}_{Q}-{\kappa}(\alpha+1)+\bar{\mathcal{V}}_d+\alpha_3+2)$ and \(\kappa=\mathcal{E}_{\mathcal{V}}-\alpha\mathcal{E}_Q\) in (\ref{c11}),
\begin{align}
   \bar{\tau}>&  \bar{\theta}(\alpha^2\mathcal{E}_{Q} + \alpha \mathcal{E}_{Q} +\bar{\mathcal{V}}_d +\alpha_3+2)+\bar{d}\nonumber\\
   &+(\mathcal{E}_{\mathcal{V}}-\alpha\mathcal{E}_Q)(\bar{\theta}(2\alpha +3)+\lambda_{max}\{K_1\}-\lambda_{min}\{K_1\})
   \label{taucon1}\\
   >& \bar{\theta}(\bar{\mathcal{V}}_d +\alpha_3+2)+\bar{d}+ \mathcal{E}_{\mathcal{V}}\omega_2-\mathcal{E}_Q\omega_3 \label{taucon2}
\end{align}
where \(\omega_2=\bar{\theta}(2\alpha +3)+\lambda_{max}\{K_1\}-\lambda_{min}\{K_1\}\), \(\omega_3=\alpha(\lambda_{max}\{K_1\}+\bar{\theta}(\alpha+2)-\lambda_{min}\{K_1\})=\alpha(\omega_2-\bar{\theta}(\alpha+1))\) and \(\delta,\omega_3>0\) since $\lambda_{max}\{K_1\}\geq\lambda_{min}\{K_1\}$. (\ref{taucon2}) implies the feasibility condition C1. Provided C1 is satisfied, since \(\bar{m}\|r(0)\|^2<\kappa_m^2\), employing Lemma 1, at $t=0$, (\ref{vdot112}) can be written as,
\begin{align}
   \dot{V}(0)<0
   \label{uub1}
\end{align}
 It can be proved from (\ref{uub1}) that $r(0^+)\in \Omega_r^{'}$ at $t=0^+$, provided $r(0)\in \Omega_r^{'}$ and (\ref{c11}) is satisfied. Repeating the above argument for all time instants,
\begin{align}
    \dot{V}(t)< 0 && \forall t \geq 0
    \label{vdot22}
\end{align}
Since $V(r,\tilde{\theta})$ in (\ref{lyap}) is positive definite, it can be inferred from (\ref{vdot22}) that
\begin{align}
 \bar{m}\|r(t)\|^2<\kappa_m^2\implies \|r(t)\|< \kappa \hspace{20pt} \forall t\geq0
  \label{lambda1}
\end{align}
Now, by solving the differential equation (\ref{fte}) and employing Assumption 2, it can be proved that 
\begin{align}
    &\|e(t)\|< \mathcal{E}_Q-\frac{\kappa}{\alpha}+\frac{\kappa}{\alpha} \hspace{22pt} \forall t\geq0\\
    &\|\dot{e}(t)\|<\alpha \mathcal{E}_Q+\kappa \hspace{15pt} \forall t\geq0
\end{align} 
Since \(\delta_1<\mathcal{E}_Q\), substituting the value of $\kappa$ (Assumption 2),

\begin{subequations}
    \begin{align}
    &\|e(t)\|<\mathcal{E}_Q \hspace{36pt} \forall t\geq0 \label{ec1}\\
    & \|\dot{e}(t)\|<\mathcal{E}_{\mathcal{V}} \hspace{36pt} \forall t\geq0
    \label{ec2}
\end{align}
\label{ec}
\end{subequations}

i.e., the trajectory tracking error and its derivative will be constrained within the user-defined safe set: $e(t), \dot{e}(t) \in \Omega_e$ $\forall t\geq0$.\\
Furthermore, since both the desired trajectory and the trajectory tracking error are bounded, i.e. $\|q_d(t)\|\leq \bar{Q}_d$, $\|e(t)\|< \mathcal{E}_Q$ and $\|\dot{e}(t)\|< \mathcal{E}_{\mathcal{V}}$, it follows directly from (\ref{ec}) that the proposed controller ensures the plant states remain within the user-defined safe set
\begin{subequations}
\begin{align}
    &\|q(t)\|<\mathcal{E}_Q+\bar{Q}_d= \bar{Q}  && \forall t \geq 0\\
     &\|\dot{q}(t)\|<\mathcal{E}_{\mathcal{V}}+\bar{\mathcal{V}}_d= \bar{\mathcal{V}}&& \forall t \geq 0
\end{align}
\end{subequations}
Since the closed-loop trajectory tracking error and the parameter estimation errors are bounded and $\theta$ is constant, it follows that the estimated parameters are also bounded, i.e., $\hat{\theta}(t)\in \mathcal{L}_{\infty}$. Consequently, the plant states $q(t), \dot{q}(t)$ and control input $\tau(t)$ remain bounded at all times. Thus, the proposed controller guarantees that all the closed-loop signals are bounded.
\end{proof}

\section{Feasibility Analysis} \label{feas}
\begin{figure}[h!]
    \centering
    \includegraphics[width=0.85\textwidth]{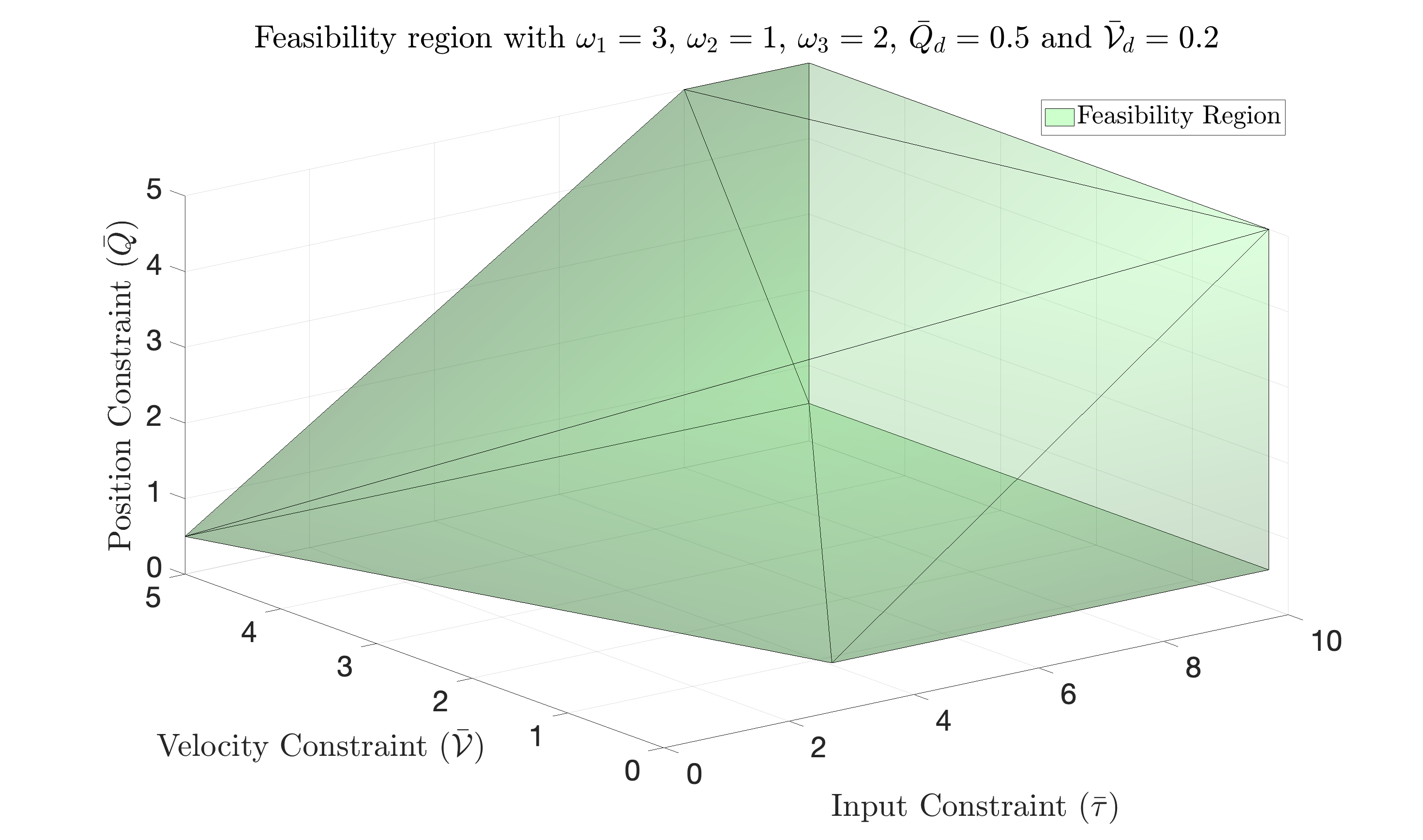}
    \caption{Caption}
    \label{feas3}
\end{figure}

Feasibility refers to the controller's ability to achieve the tracking objective while adhering to user-defined state and input constraints in the presence of disturbances. The feasibility condition \( C1 \) establishes a lower bound on the input constraint, emphasizing the trade-off between the state and input constraints and the upper bound of the disturbance. Fig. \ref{feas3} shows the relationship and the associated trade-off between the position, velocity, and input constraints.

\begin{figure}[h!]
     \centering
     \begin{subfigure}[b]{0.4\textwidth}
         \centering
         \includegraphics[width=\textwidth]{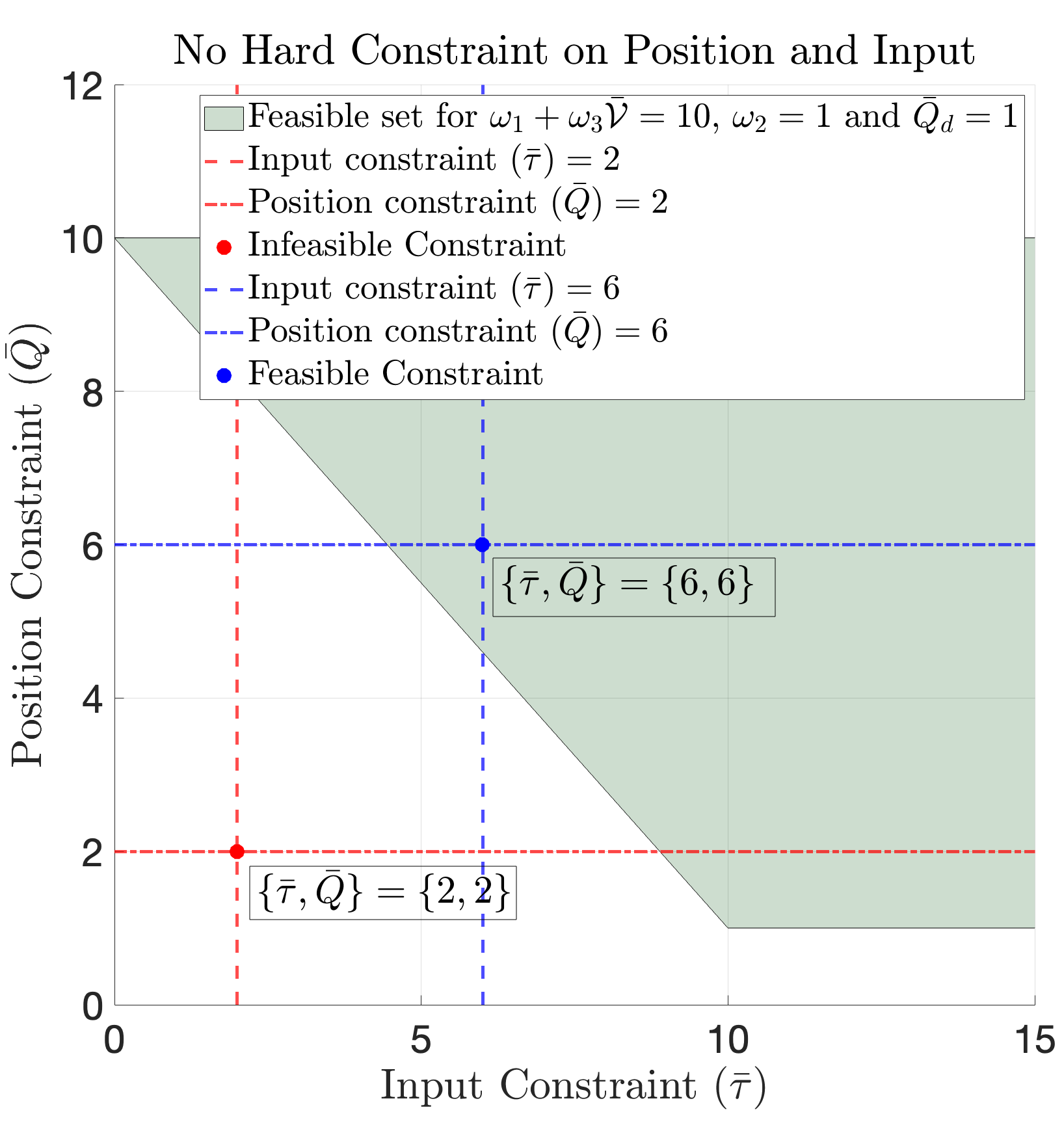}
         \caption{}
         \label{fno_hardcon}
     \end{subfigure}
     ~
     \begin{subfigure}[b]{0.4\textwidth}
         \centering
         \includegraphics[width=\textwidth]{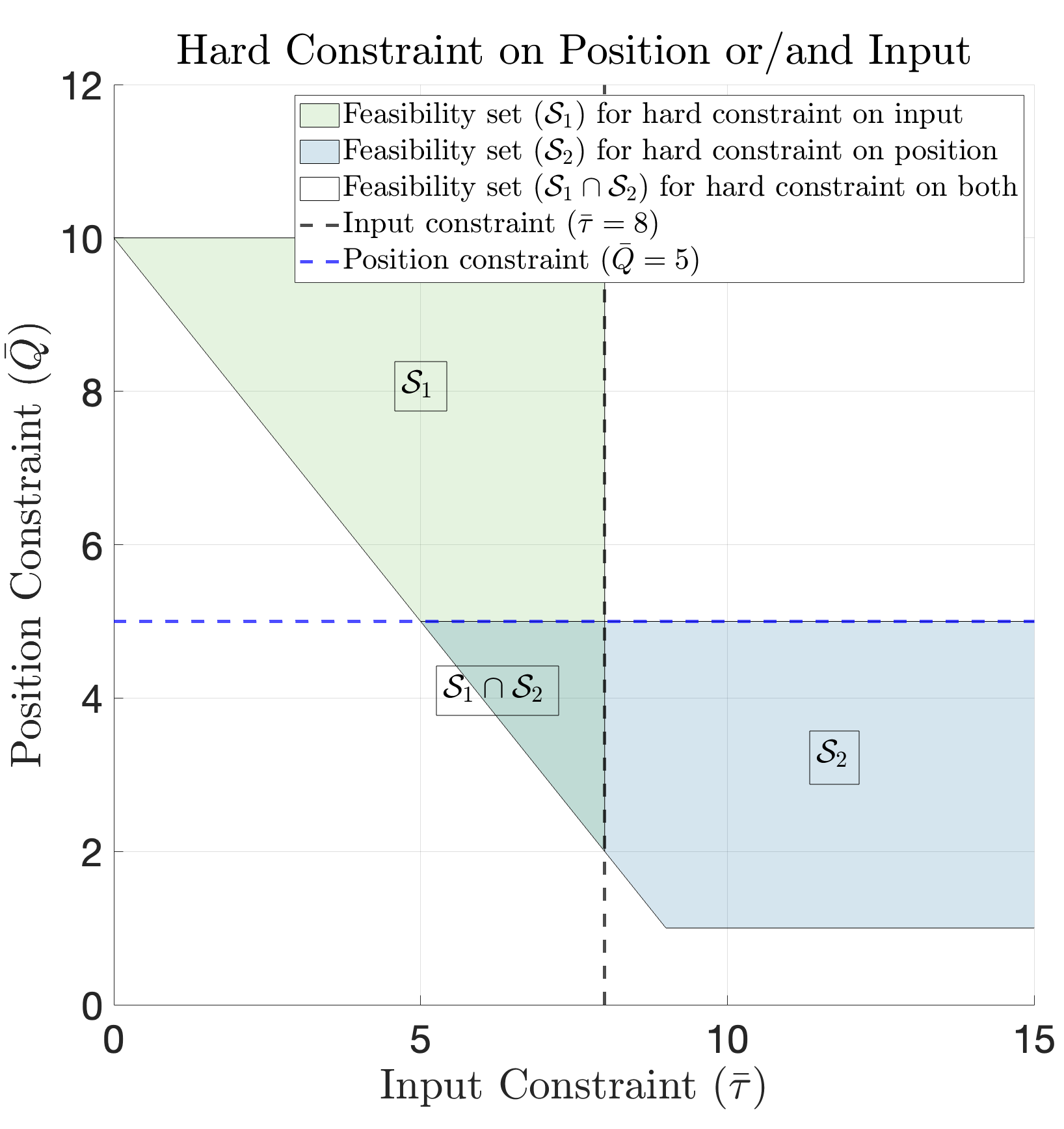}
         \caption{}
         \label{Hardcon_both}
     \end{subfigure}
     \caption{Feasibility regions for Case 1}
     \label{f1f2}
     \end{figure}

\begin{figure}[h!]
     \begin{subfigure}[b]{0.4\textwidth}
         \centering
         \includegraphics[width=\textwidth]{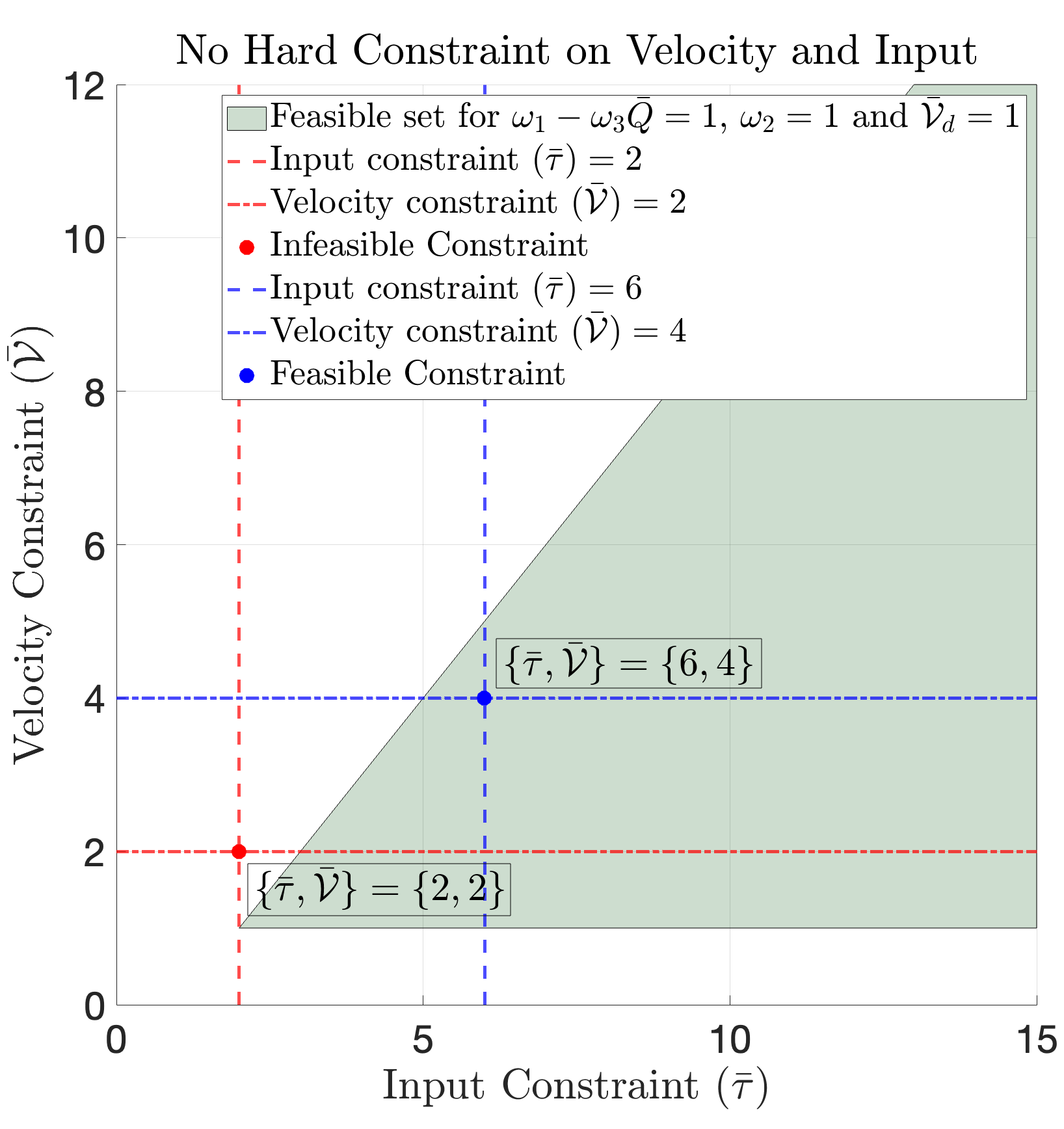}
         \caption{}
         \label{fno_hardcon_2}
     \end{subfigure}
     ~
     \begin{subfigure}[b]{0.4\textwidth}
         \centering
         \includegraphics[width=\textwidth]{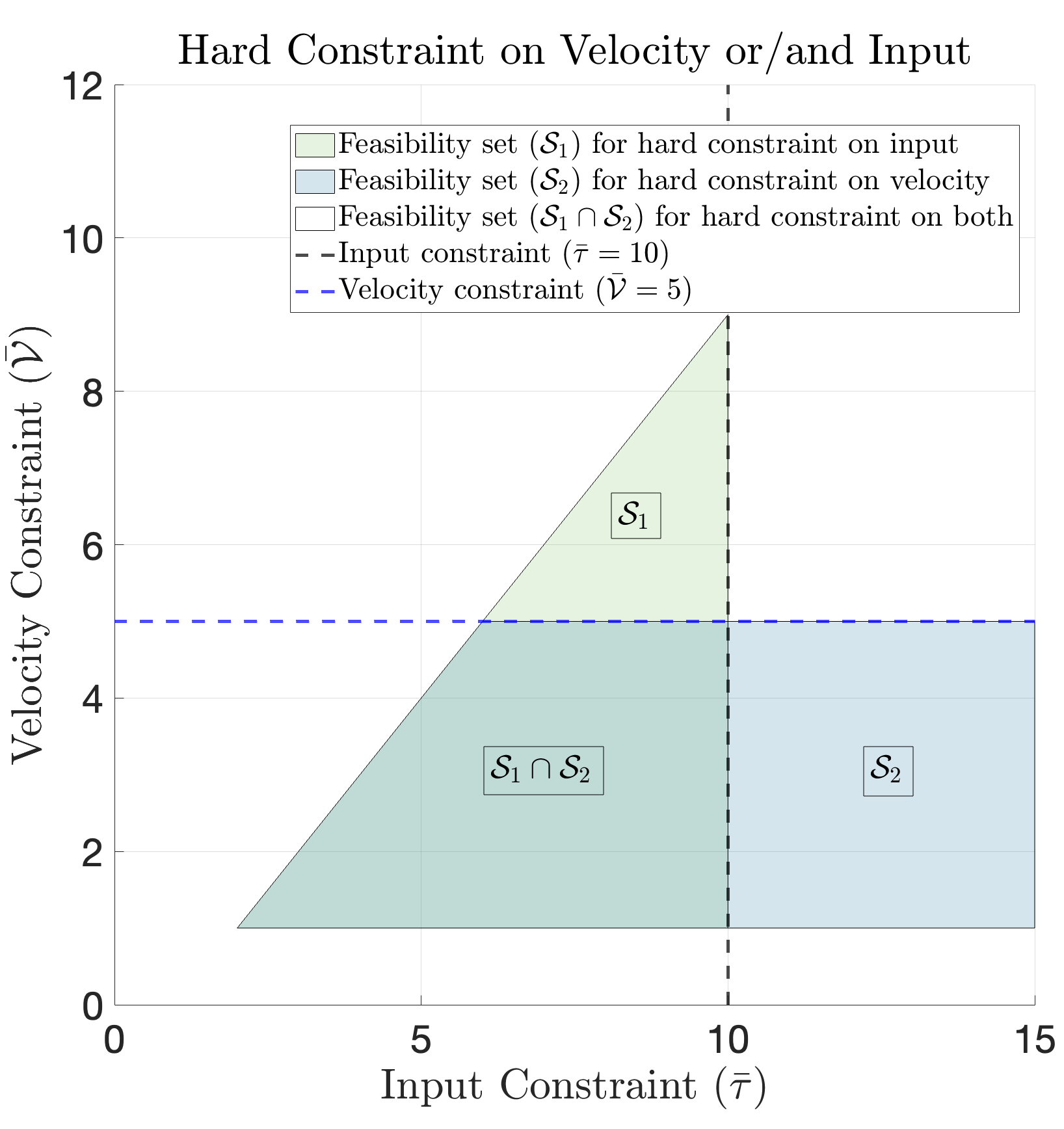}
         \caption{}
         \label{Hardcon_both_2}
     \end{subfigure}
      \caption{Feasibility region for Case 2}
     \label{f3f4}
     \end{figure}

\begin{figure}[h!]
     \begin{subfigure}[b]{0.4\textwidth}
         \centering
         \includegraphics[width=\textwidth]{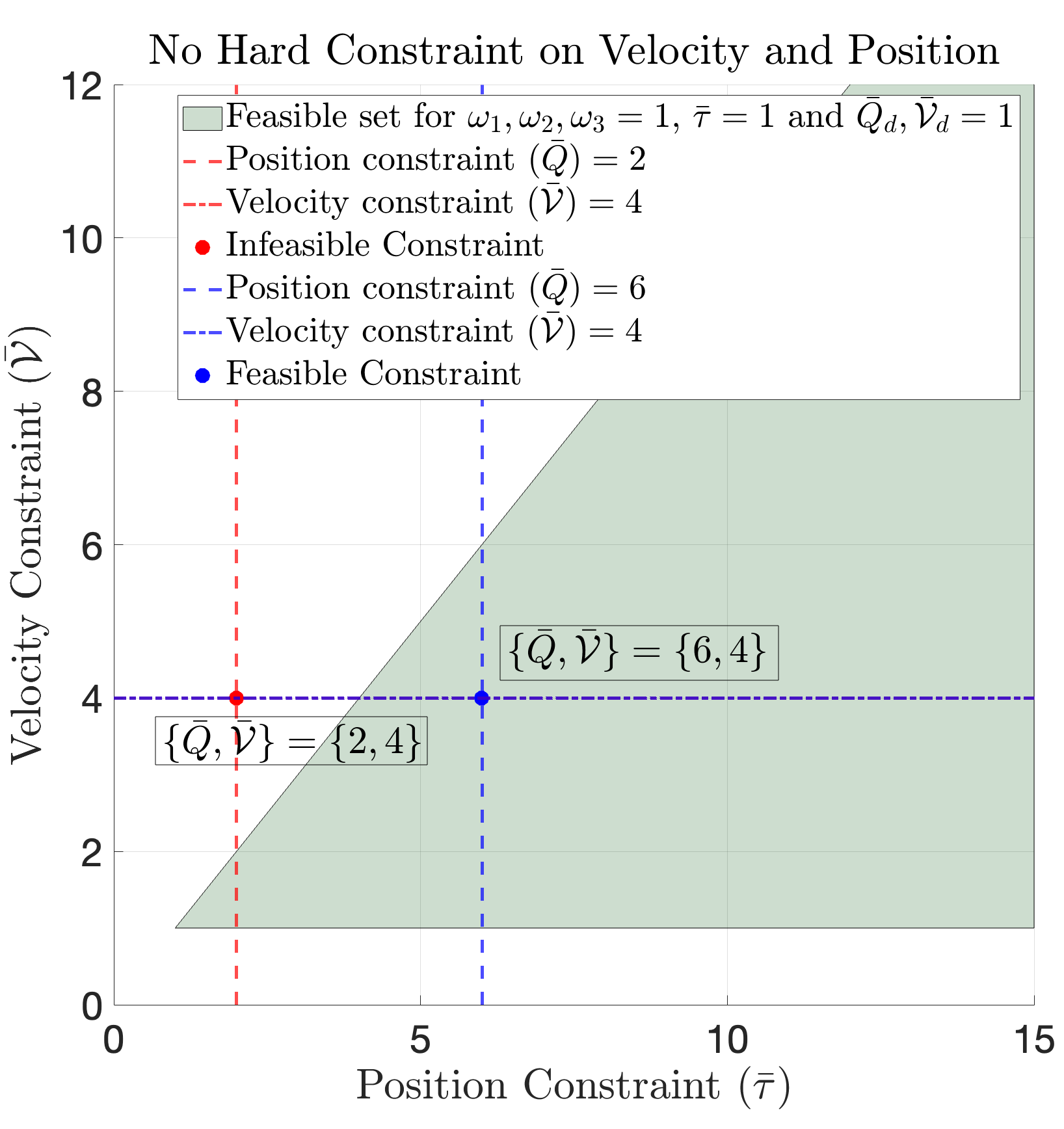}
         \caption{}
         \label{fno_hardcon_3}
     \end{subfigure}
     ~
     \begin{subfigure}[b]{0.4\textwidth}
         \centering
         \includegraphics[width=\textwidth]{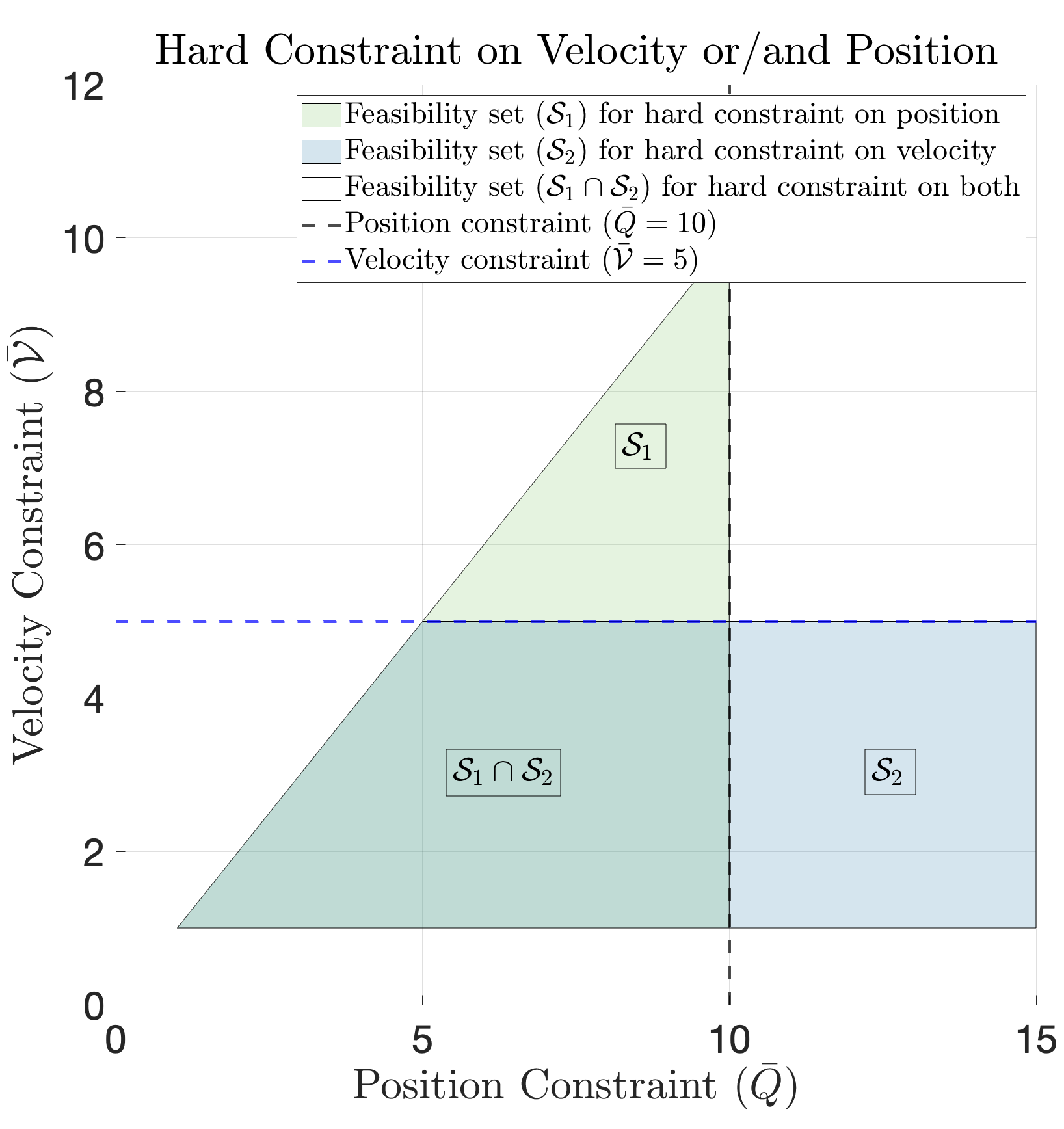}
         \caption{}
         \label{Hardcon_both_3}
     \end{subfigure}
     \caption{Feasibility region for Case 3}
     \label{f5f6}
     \end{figure}

The evaluation of the controller's feasibility can be approached in the following ways:\\~\\
\textbf{Case 1: {Trade-off between $\bar{\tau}$ and $\bar{Q}$ for a fixed $\bar{\mathcal{V}}$}} \\
\textit{1.1. No hard constraints on $\bar{\tau}$ and $\bar{Q}$:}
If the user does not provide any hard constraints on the position or the control input, the feasibility region in Fig. (\ref{fno_hardcon}), obtained from the feasibility condition C1, shows the possible values of the state and input constraints that result in a feasible control policy, e.g., in Fig. (\ref{fno_hardcon}), the constraints $\{ \bar{\tau},\bar{Q}\}=\{2,2\}$ are not feasible, while $\{ \bar{\tau}, \bar{Q}\}=\{6,6\}$ are feasible.\\
\textit{1.2. Hard constraint on the $\bar{\tau}$ or/and  $\bar{Q}$:} Given a position constraint $(\bar{Q})$, the lower bound for the input constraint $\bar{\tau}$ can be determined using the feasibility condition C1. Conversely, with a hard input constraint $\bar{\tau}$, C1 provides the minimum allowable constraint on the position $(\bar{Q})$. Fig. \ref{Hardcon_both} illustrates the feasibility sets for imposing hard constraints on the input $(\mathcal{S}_1)$, the position $(\mathcal{S}_2)$, and both the input and position $(\mathcal{S}_1 \cap \mathcal{S}_2)$. As shown in Fig. \ref{Hardcon_both}, when $\bar{Q} = 5$, the minimum allowable $\bar{\tau}$ is 5. Tightening $\bar{{Q}}$ requires increasing $\bar{\tau}$. Similarly, with $\bar{\tau} = 8$, the minimum allowable $\bar{{Q}}$ is 2.
\\
\textbf{Case 2: {Trade-off between $\bar{\tau}$ and $\bar{\mathcal{V}}$ for fixed $\bar{Q}$}} \\
\textit{2.1. No hard constraints on $\bar{\tau}$ and $\bar{\mathcal{V}}$:} For any given set of $\{ \bar{\tau},\bar{\mathcal{V}}\}$, feasibility can be verified using C1. As shown in Fig. (\ref{fno_hardcon_2}), $\{ \bar{\tau},\bar{\mathcal{V}}\}=\{2,2\}$ is infeasible; however, $\{ \bar{\tau}, \bar{\mathcal{V}}\}=\{6,4\}$ is feasible.\\
\textit{2.2. Hard constraint on the $\bar{\tau}$ or/and  $\bar{\mathcal{V}}$:} Given a $(\bar{\mathcal{V}})$, the minimum required $\bar{\tau}$ can be determined using C1 and vice versa. Fig. \ref{Hardcon_both} illustrates the feasibility sets, i.e., when $\bar{\mathcal{V}} = 5$, the minimum  $\bar{\tau}$ is 2. Increasing $\bar{\mathcal{V}}$ requires a larger $\bar{\tau}$ and with $\bar{\tau} = 10$, the minimum allowable $\bar{\mathcal{V}}$ is 1.\\
\textbf{Case 3: {Trade-off between $\bar{Q}$ and $\bar{\mathcal{V}}$ for fixed $\bar{\tau}$}} \\
\textit{3.1. No hard constraints on $\bar{Q}$ and $\bar{\mathcal{V}}$:} For any given $\{ \bar{Q},\bar{\mathcal{V}}\}$, feasibility is verified using C1. In Fig. (\ref{fno_hardcon_3}),  $\{ \bar{\tau}, \bar{\mathcal{V}}\}=\{6,4\}$ is feasible, while $\{ \bar{Q},\bar{\mathcal{V}}\}=\{2,4\}$ is infeasible.\\
\textit{3.2. Hard constraint on the $\bar{Q}$ or/and  $\bar{\mathcal{V}}$:} Given a $\bar{Q}$, the maximum allowable $\bar{\mathcal{V}}$ can be obtained using C1. The minimum value of $\bar{\mathcal{V}}$ will depend on $\bar{\mathcal{V}}_d$ (Assumption 1). Fig. (\ref{Hardcon_both_3}) depicts the feasibility sets for hard constraint on position $(\mathcal{S}_1)$, velocity $(\mathcal{S}_2)$, and both $(\mathcal{S}_1 \cap \mathcal{S}_2)$. When $\bar{\mathcal{V}} = 5$, the minimum allowable $\bar{Q}$ is 1. Increasing $\bar{\mathcal{V}}$ demands a larger $\bar{Q}$. Similarly, with $\bar{Q} = 10$, the feasible range of $\bar{\mathcal{V}}$ is $
[1,10]$.
\begin{remark}
The feasibility condition shows a fundamental trade-off between position, velocity, and input constraints. For a fixed velocity constraint, increasing the position constraint enables more flexible state regulation, reducing the required control input. Conversely, for a fixed position constraint, increasing the velocity constraint demands a higher control authority to ensure tracking to a desired reference trajectory. Similarly, when the input constraint is fixed, ensuring feasibility under a tighter velocity constraint requires the corresponding relaxation of the position constraint.
\end{remark}

\section{Simulation Results}
\begin{figure}[h!]
\centering
\includegraphics[width=0.9\linewidth]{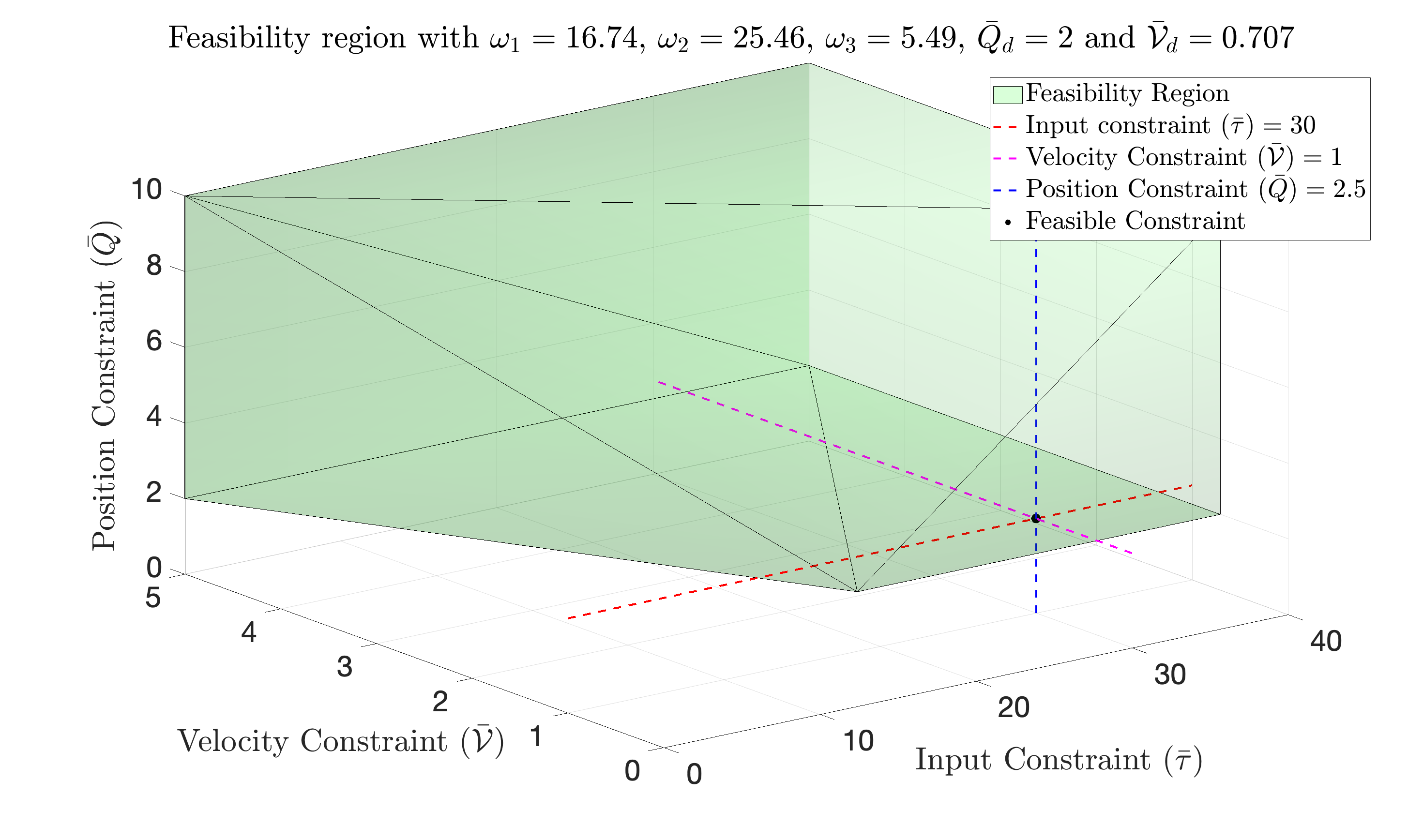}
\caption{Feasibility region for (\ref{plant}).}
\label{f3}
\end{figure}

To illustrate the effectiveness of the proposed algorithm for constrained and uncertain Euler-Lagrange (E-L) systems, we consider the dynamics of a two-link robotic manipulator.
\begin{equation}
M(q)\ddot{q}+V_m(q,\dot{q})\dot{q}+F_d(\dot{q})+G_r(q)=\tau +d  
\label{plant}
\end{equation}
where, $q(t)=[q_1(t), q_2(t)]^T\in \mathbb{R}^2$ denotes the angular position (rad) and $\dot{q}(t)=[\dot{q}_1(t), \dot{q}_2(t)]^T\in \mathbb{R}^2$ represents the angular velocity of the manipulator. The matrices associated with the plant (\ref{plant}) are given by:
\begin{align*}
    &M(q)=\begin{bmatrix}
    p_1+2p_3c_2 & p_2+p_3c_2\\
    p_2+p_3c_2 & p_2
    \end{bmatrix}\\
    &V_m(q,\dot{q})=\begin{bmatrix}
    -p_3s_2\dot{q_2} & -p_3s_2(\dot{q_1}+\dot{q_2})\\
    p_3s_2\dot{q_1}  & 0
    \end{bmatrix}\\
    &F_d(\dot{q})=\begin{bmatrix}
    f_{d_1}\\
    f_{d_2}
    \end{bmatrix}\\
    &G_r(q)=0_{2\times 1}
\end{align*}
The desired trajectory is given by
\begin{equation}
   q_d(t)=\begin{bmatrix}
    0.5sint \\ 2cos(t/4)
    \end{bmatrix}
\end{equation}
The objective is to design an adaptive tracking controller such that plant states $q(t)$ and $\dot{q}(t)$  track the desired reference trajectories $q_d(t)$ and $\dot{q}_d(t)$, respectively, while simultaneously adhering to user-defined constraints on state and input. The constraints are chosen such that the feasibility condition C1 and the gain condition,  are satisfied. In this example, we consider hard constraints on the states,given by 
\begin{align*}
    &\|q(t)\|<\bar{Q}=2.5\\
    &\|\dot{q}(t)\|<\bar{\mathcal{V}}=1
\end{align*}
The other parameters used for simulation are chosen as: $\Gamma=10\mathbb{I}_{m\times m}$, , 
$\bar{Q}_d=2$, $\bar{\mathcal{V}}_d=0.707$, $\alpha_3=0.3$, $\bar{d}=5$, $K_1=\begin{bmatrix}
    1.5 & 0\\
    0 & 1
\end{bmatrix}$, $\bar{\theta}=6.2$, $p_1=3.473$ kg-m, $p_2=0.196$ kg-m, $p_3=0.242$ kg-m, $f_{d_1}=5.3$ N s, $f_{d_2}=1.1$ N s. The bounded external disturbance is considered as 
\begin{align}
   d(t)=\begin{cases}
       \begin{bmatrix}
           0 & 0
       \end{bmatrix}^T & \text{if} \hspace{6pt} t<100\\
       \begin{bmatrix}
           3\sin(t) & 3\cos(t)
       \end{bmatrix}^T & \text{if} \hspace{6pt} 100\leq t < 200\\
       \begin{bmatrix}
          5\sin(t) & 5\cos(t) 
       \end{bmatrix}^T & \text{if} \hspace{6pt} 200\leq t < 300\\
   \end{cases} 
   \label{dis_function}
\end{align}
Given $\|q_d\| \leq 2$, $\|\dot{q}_d\| \leq 0.7$, the state constraint can be transformed to the constraint on the trajectory tracking errors, i.e. $\|e\|<\mathcal{E}_Q=0.5$, $\|\dot{e}\|<\mathcal{E}_{\mathcal{V}}=0.293$. Using the gain condition in , \(\alpha<0.58\) and for the simulation, we choose \(\alpha=0.5\). Since $\kappa=\mathcal{E}_{\mathcal{V}}-\alpha \mathcal{E}_Q=0.04$, from the proof of Theorem 1, it can be easily inferred that satisfying the constraint on the filtered tracking error $\|r\|<0.04$ will consequently ensure the satisfaction of the state constraint. Using (\ref{taucon1}), the minimum value of the input constraint is calculated as \(\bar{\tau}>28.5\) and for simulation we have chosen \(\bar{\tau}=30\). \\
Fig. \ref{f3} illustrates the feasibility region with $\bar{Q}=2.5$, $\bar{\mathcal{V}}=1$ and $\bar{\tau}=30$ satisfying C1. Since $\bar{Q}_d=2$ and $\bar{\mathcal{V}}_d=0.707$, the minimum feasible position and velocity constraints are $\bar{Q}_{min}>2$ and $\bar{\mathcal{V}}_{min}>0.707$, respectively. Within the feasibility region, any combination of input and state constraints ($\{\bar{\tau},\bar{Q}, \bar{\mathcal{V}}\}$) ensures the existence of a feasible control policy.

\begin{figure}[h!]
\centering
\includegraphics[width=0.95\linewidth]{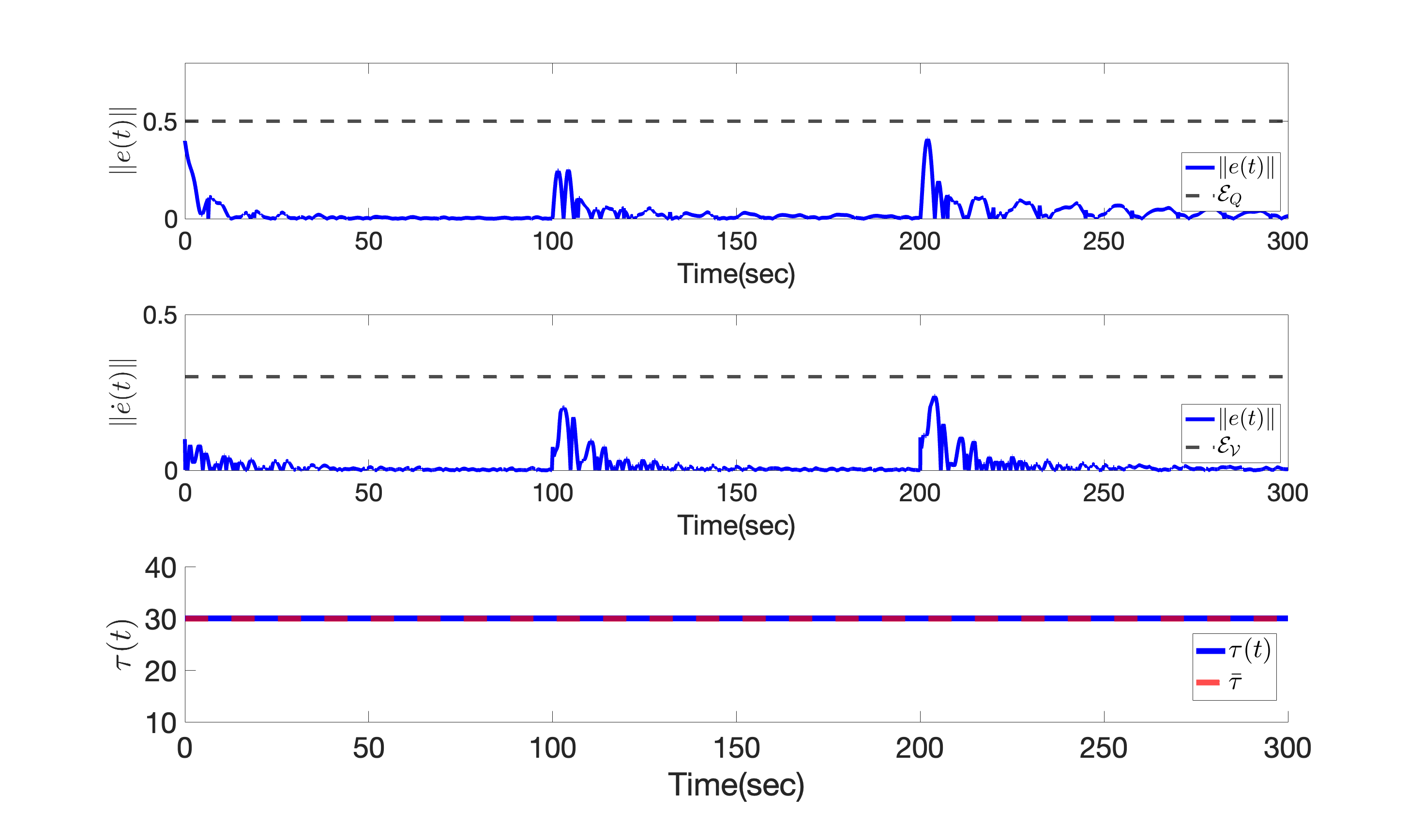}
\caption{Trajectory tracking errors between the actual and desired angular positions and velocities and corresponding control input using the proposed controller in the presence of bounded external disturbances (\ref{dis_function}).}
\label{E_tau_plot}
\end{figure}

\begin{figure}[h!]
\centering
\includegraphics[width=0.95\linewidth]{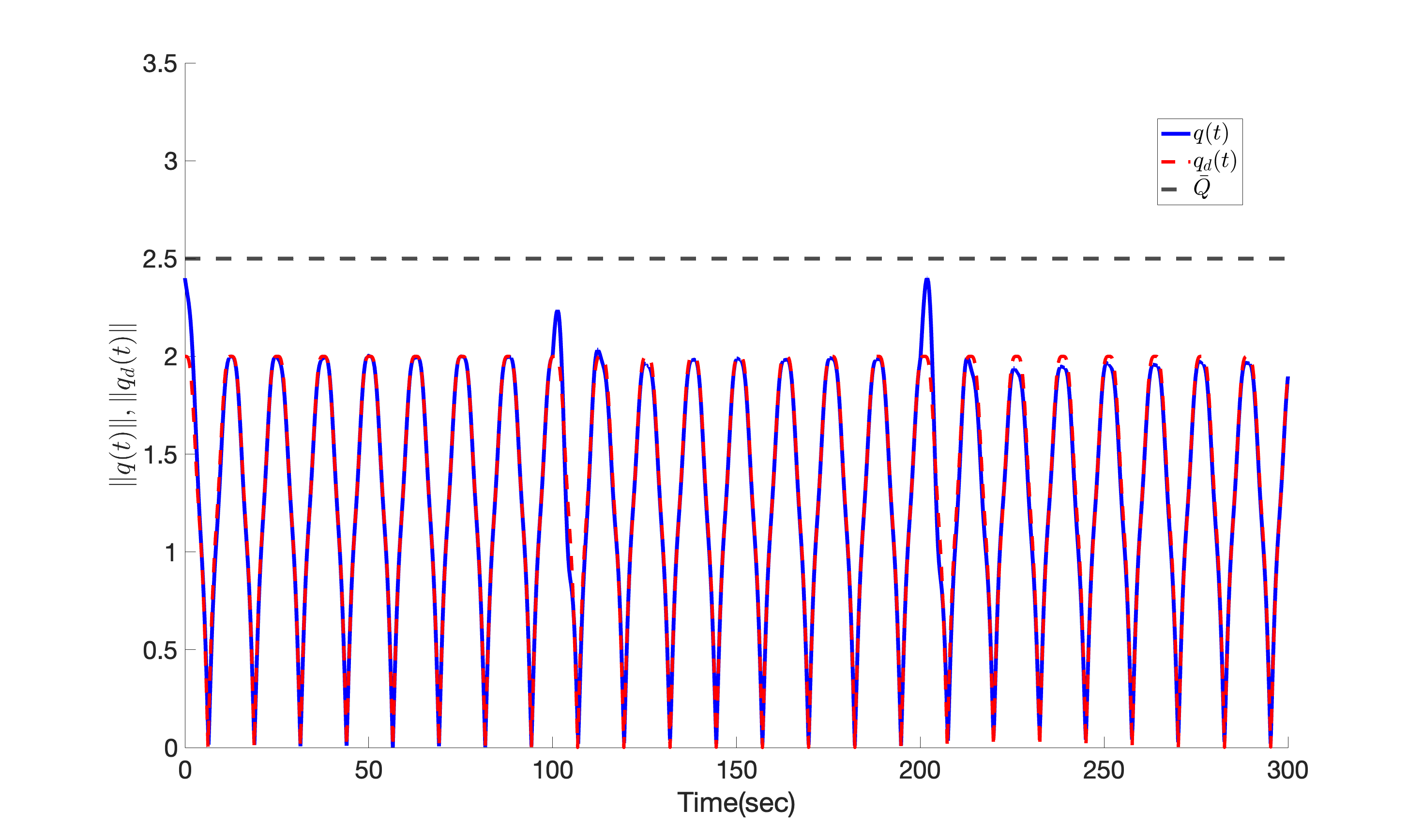}
\caption{Tracking performance of the actual position to the desired position using the proposed controller in the presence of bounded external disturbances (\ref{dis_function}).}
\label{qplot}
\end{figure}

\begin{figure}[h!]
\centering
\includegraphics[width=0.95\linewidth]{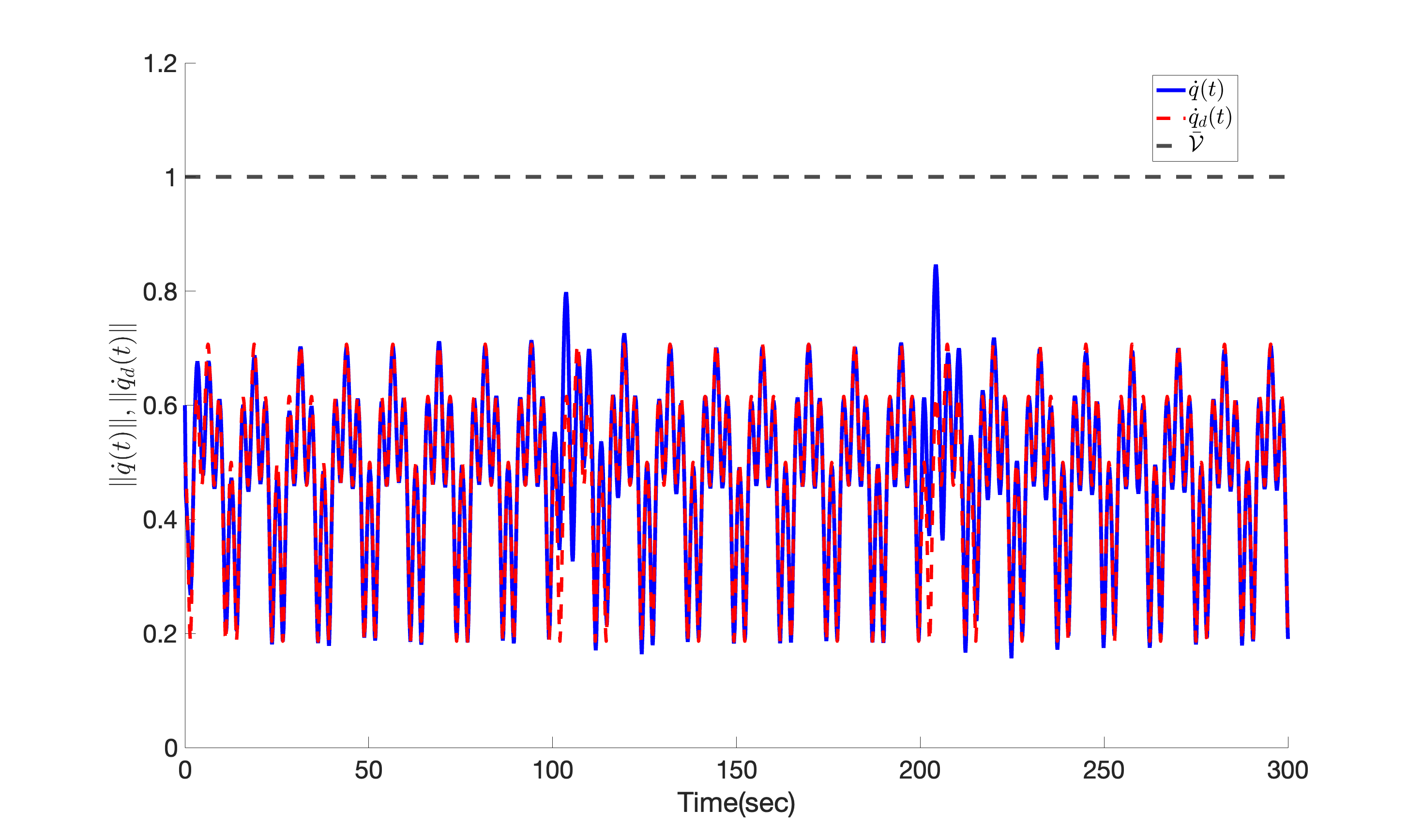}
\caption{Tracking performance of the actual velocity to the desired velocity using proposed controller in presence of bounded external disturbances (\ref{dis_function}).}
\label{qdotplot}
\end{figure}

 The proposed controller guarantees that the trajectory tracking errors are within the user-defined safe set (Fig. \ref{E_tau_plot}), which in turn implies that the plant states, i.e., the position ($\|q(t)\|$) and velocity ($\|\dot{q}(t)\|$) of the manipulator remain within the pre-specified limit (Fig. \ref{qplot}-\ref{qdotplot}), while tracking the desired reference trajectory.  Furthermore, Fig. \ref{E_tau_plot} shows that the required control input is within the pre-defined safe set, i.e. $\|\tau(t)\|<30$. The disturbance (\ref{dis_function}) is injected at $t=100$ sec and $t=200$ sec, and the proposed controller ensures state and input constraints satisfaction even in the presence of disturbances. Note that the set of feasible constraints has been chosen considering $\bar{d}=5$, leading to a slightly conservative result for $t<200$ sec.

To show the effectiveness of the proposed control law, we compare it with a robust adaptive \cite{arc1} controller where the adaptive gain is chosen as $\Gamma_c=20\mathbb{I}_{m\times m}$. Note that adaptation gains for both the proposed controller and the classical method are tuned to achieve comparable tracking performance.

\begin{figure}[h!]
\centering
\includegraphics[width=0.95\linewidth]{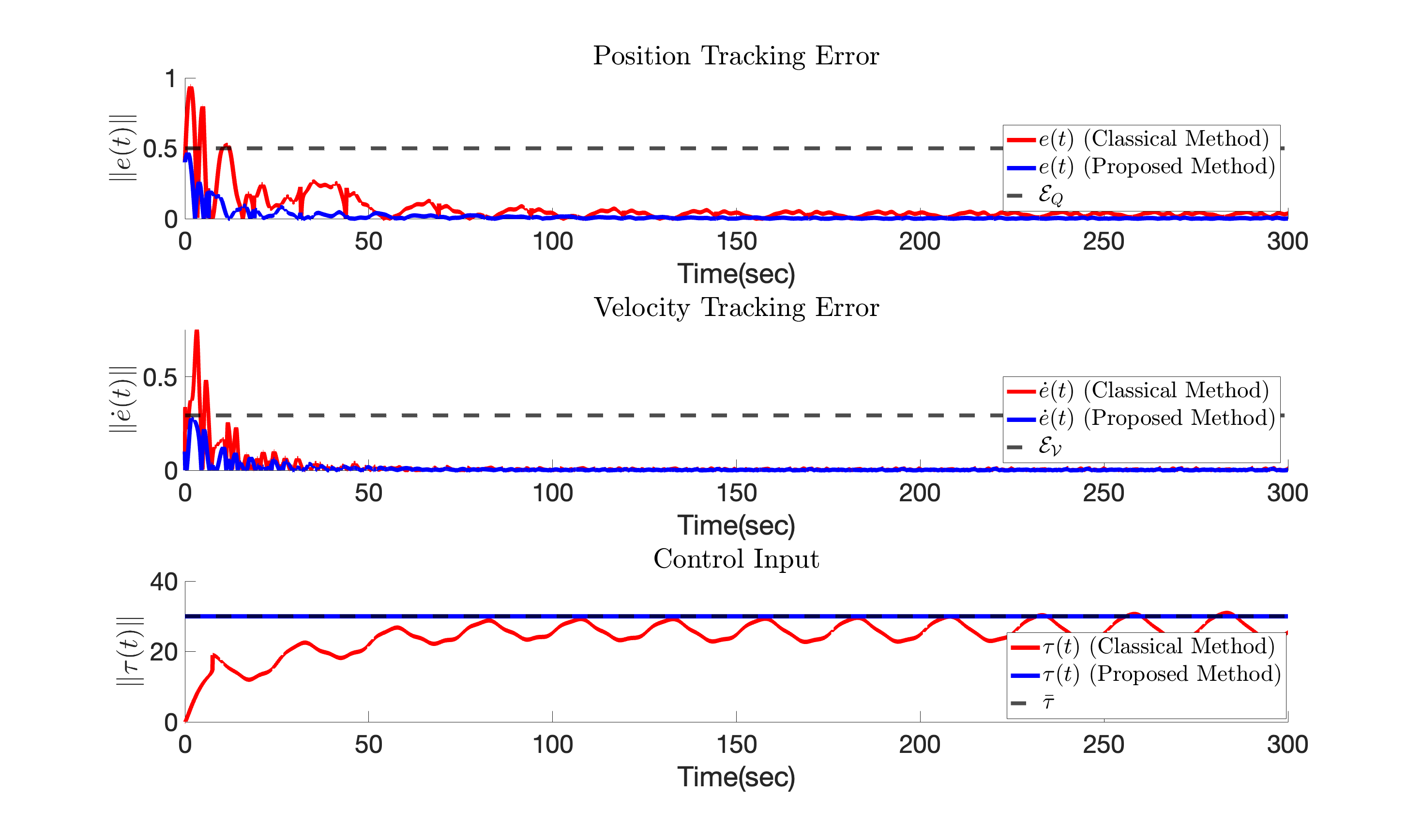}
\caption{Comparison of trajectory tracking errors and required control input using the proposed controller and a robust adaptive tracking controller in the presence of bounded external disturbances ($\bar{d}=5$).}
\label{qplot_com}
\end{figure}

Fig. \ref{qplot_com} shows that the proposed controller ensures the trajectory tracking errors remain within the user-defined safe set; however, tracking errors go beyond the safe region using the classical method. Additionally, the proposed control technique keeps the control effort within the user-defined constrained bounds, while the classical approach violates the input constraints. For comparative analysis, we have considered $d(t)=\begin{bmatrix}
    5\sin(t) & 5\cos(t)
\end{bmatrix}^T$, $\forall t\geq 0$ and $\bar{d}=5$.

The tuning parameter $\alpha$ directly affects the filtered tracking error, balancing position error $e(t)$ and velocity error $\dot{e}(t)$. A large value of $\alpha$ reduces $e(t)$ faster, making it preferable for applications requiring precise positioning, while a smaller \(\alpha\) helps minimizing the velocity error $\dot{e}(t)$, which is beneficial for high-speed tracking. In classical methods, where no bounds are imposed on $\alpha$, its value can be arbitrarily large or small, leading to trade-offs where either the position or velocity constraints may be violated. However, the proposed method imposes an explicit upper bound on \(\alpha\) (\ref{alpha}), ensuring the satisfaction of both constraints.\\
With a classical adaptive controller, satisfactory tracking performance may be achieved by tuning the adaptive gains, albeit through an ad hoc and trial-and-error gain tuning process. However, increasing the adaptive gain to improve tracking comes with trade-offs: it significantly amplifies the control effort, which may violate the user-defined constraint, and introduces oscillatory behavior in the system response, compromising stability and overall performance. While it may be possible to satisfy both state and input constraints in classical adaptive control by carefully tuning the adaptation gains, this often comes at the cost of degraded tracking performance. This inherent trade-off between performance and constraint satisfaction is a common limitation of classical adaptive controllers. In contrast, the proposed controller effectively addresses these challenges by ensuring tracking to desired reference trajectories while simultaneously satisfying pre-specified bounds on plant states and control input without any ad hoc gain tuning. This capability increases the proposed controller's suitability for a wide range of applications in safety-critical systems.

\section{Conclusion}
In this paper, we design an adaptive tracking control algorithm for uncertain E-L systems with state and input constraints that ensures tracking in the presence of bounded external disturbances. The control strategy integrates a Barrier Lyapunov Function (BLF)-based approach with a saturated controller, ensuring that both plant states and control inputs remain within predefined safe boundaries as the system tracks a desired reference trajectory. The proposed controller also guarantees that the trajectory tracking error remains constrained within a known limit and all the closed-loop signals remain bounded. Furthermore, the paper provides verifiable conditions to guarantee the feasibility of the control policy. Simulation results illustrate the advantages of the proposed control scheme, showing its effectiveness and improved performance over traditional robust adaptive controllers.

\bibliographystyle{ieeetr}
\bibliography{ref}
\end{document}